\documentclass[twocolumn, preprintnumbers, superscriptaddress,
  prl]{revtex4-1}
\usepackage{amssymb,amsmath}
\usepackage{graphicx}

\begin{document}
\preprint{INR-TH-2016-034}
\title{Relativistic axions from collapsing Bose stars}

\author{D.G. Levkov}\email{levkov@ms2.inr.ac.ru}
\affiliation{Institute for Nuclear Research of the Russian Academy
  of Sciences, 60th October Anniversary prospect 7a, Moscow 117312,
  Russia}
\author{A.G. Panin}%\email{panin@ms2.inr.ac.ru}
\affiliation{Institute for Nuclear Research of the Russian Academy
  of Sciences, 60th October Anniversary prospect 7a, Moscow 117312,
  Russia}
\affiliation{Moscow Institute of Physics and Technology, 141700,
  Dolgoprudny, Russia}
\author{I.I. Tkachev}%\email{tkachev@ms2.inr.ac.ru}
\affiliation{Institute for Nuclear Research of the Russian Academy
  of Sciences, 60th October Anniversary prospect 7a, Moscow 117312,
  Russia}

\begin{abstract}
The substructures of light bosonic (axion-like) dark matter may
condense into compact Bose stars. We study collapses of the
critical-mass stars caused by attractive self-interaction of the
axion-like particles and find that these processes proceed in an
unexpected universal way. First, nonlinear self-similar evolution
(called ``wave collapse'' in condensed matter physics) forces the particles
to fall into the star center. Second, interactions in the dense center
create an outgoing stream of mildly relativistic particles
which carries away an essential part of the star mass. The collapse
stops when the star remnant is no longer able to support the
self-similar infall feeding the collisions. We shortly discuss possible
astrophysical and cosmological implications of these phenomena.  
\end{abstract}

\maketitle

%%%%%%%%%%%%%%%%%%%%%%%%%%%%%%%%%%%%
\paragraph{1. Introduction.}
Increasingly stringent experimental constraints~\cite{Ulmer:2016ljv}
on low-energy supersymmetry reignited discussion of
non-supersymmetric dark matter candidates such as  the QCD
axion~\cite{Kim:2008hd} and axion-like particles
(ALP)~\cite{Arvanitaki:2009fg}. The interest is heated up by fast
progress~\cite{Ringwald:2012hr} in ALP searches and peculiar
properties of the axion-like dark matter related, in particular, to
the misalignment mechanism of its generation~\cite{Preskill:1982cy},
see also~\cite{Kolb:1993zz,Hiramatsu:2012gg}.

A very special possibility opening up due to tiny velocities and
large occupation numbers of the ALP dark matter is formation of
Bose stars~\cite{Ruffini:1969qy, Chavanis:2011, Eby:2016ab}~---
gravitationally bound puddles of the ALP Bose condensate. These
objects were observed as ``solitonic galaxy cores'' in numerical 
simulations~\cite{Schive:2014dra}  of structure formation by ultralight
($m \sim 10^{-22} \, \mathrm{eV}$) ALP dark matter. At larger ALP
masses and, notably, in the case of the QCD axion, the Bose stars were
conjectured~\cite{Kolb:1993zz, Seidel:1993zk} to appear in the centers
of the axion
miniclusters~\cite{Kolb:1994fi, Hardy:2016mns} which populate the
Universe if the Peccei-Quinn symmetry is broken after
inflation, cf.~\cite{Guth:2014hsa}. Although the last mechanism
is not yet confirmed by numerical modeling, it is safe to say that the
axion Bose stars constitute a part of the dark mass at least in some
ALP dark matter models.

Rich phenomenology of the Bose stars is related to the peculiar
self-interaction potentials~\cite{Kim:2008hd, Arvanitaki:2009fg} of
the ALP and QCD axion fields $a(x)$, 
\begin{equation}
  \label{eq:2}
  {\cal V} = m^2 f_a^2\left( \theta^2 /2 -  g_4^2\theta^4/4! + 
  \dots\right)\;,\qquad  \theta\equiv a/f_a\;,
\end{equation}
which contain, besides the mass $m$, interaction terms
suppressed by the new physics scale $f_a$. Typically, ${\cal
  V}(\theta)$ is periodic and the quartic constant $\lambda_4 \equiv - 
g_4^2 m^2/f_a^2$ is negative. This introduces attraction between the
ALP and causes~\cite{Chavanis:2011} collapse of 
large-mass Bose stars which previously was studied
in~\cite{Chavanis:2016, Eby:2016} using the Gaussian Ansatz.

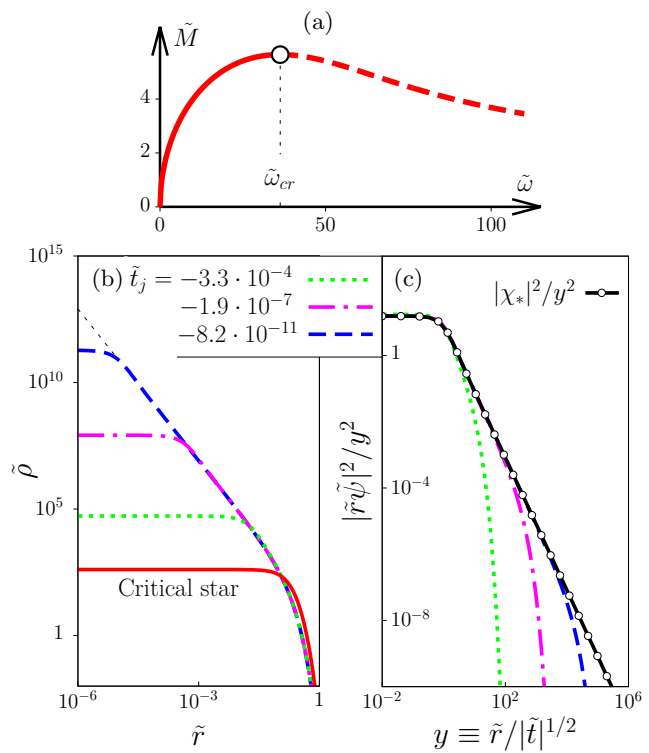
\begin{figure}[t]
\unitlength=0.01\columnwidth
\begin{picture}(95,115)
%\put(8,78){\includegraphics[height=0.31\columnwidth]{fig_collapse.pdf}}
\put(17,78){\includegraphics[height=0.34\columnwidth]{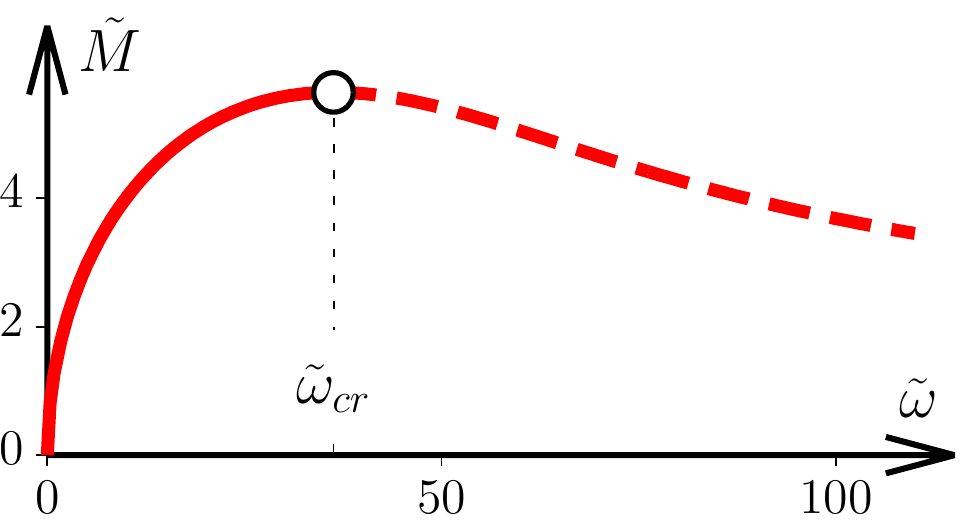}}
\put(-3,-0.4){\includegraphics[height=0.775\columnwidth]{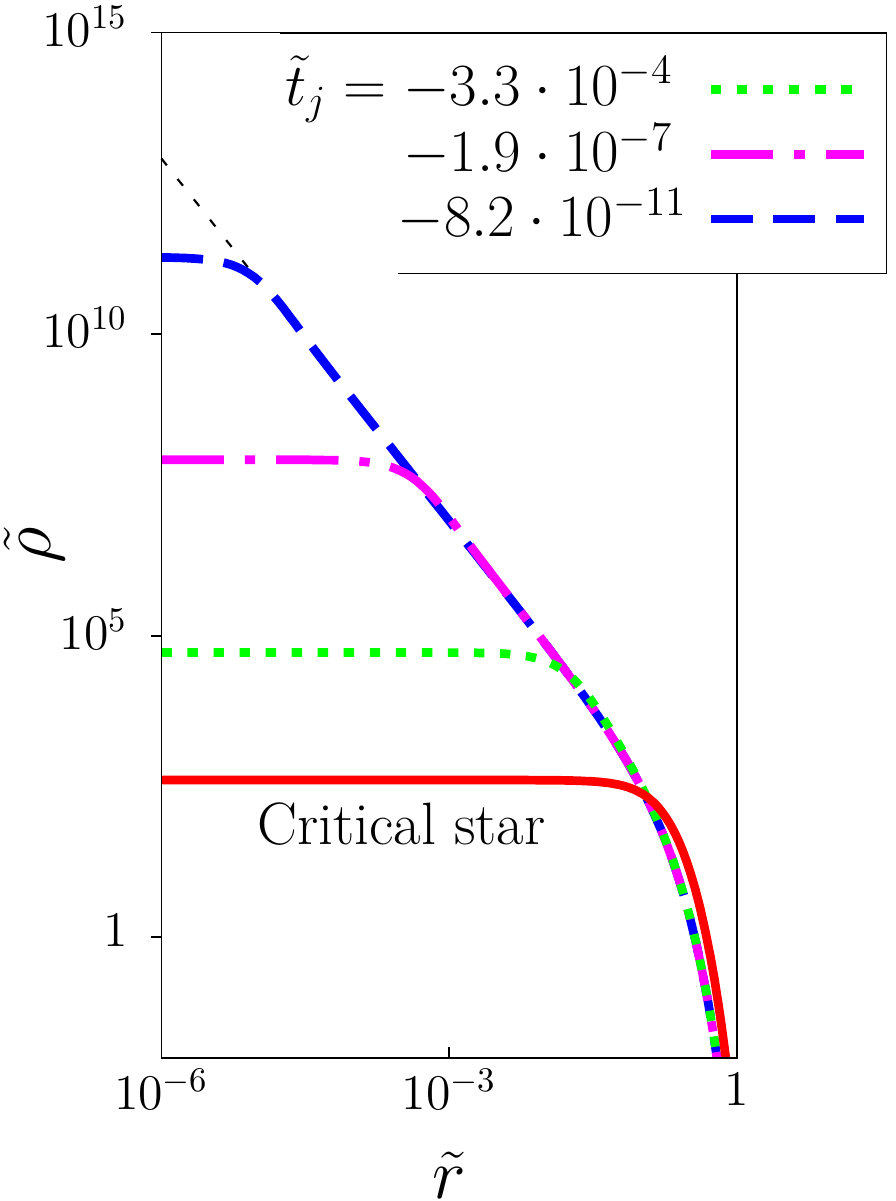}}
\put(46.1,-1.2){\includegraphics[height=0.763\columnwidth]{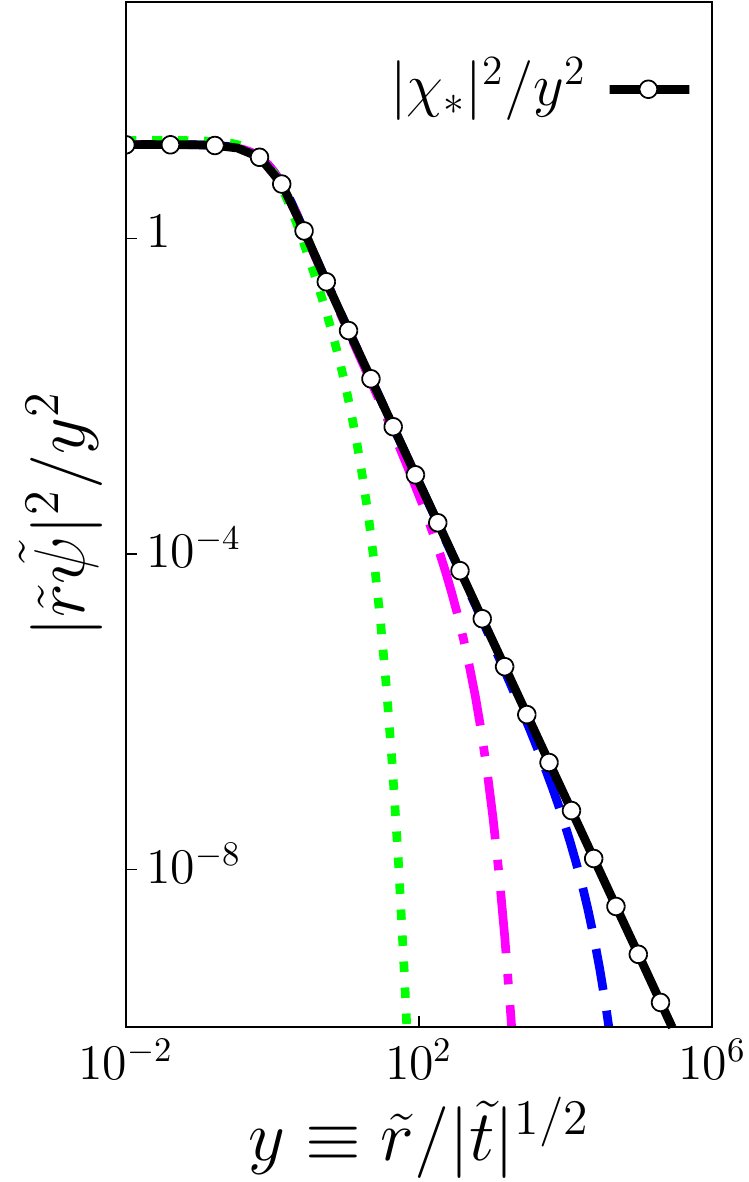}}
%\put(0,8.8){\line(1,0){95}}
%\put(0,75){\line(1,0){95}}
%\put(0,114){\line(1,0){95}}
%\put(18.7,112){(a)}
%\put(21.7,0){\line(0,1){120}}
\put(9.5,70.5){(b)}
\put(42,110){(a)}
\put(56,70.5){(c)}
\end{picture}
\caption{(a)~Star mass as a function of the binding
  energy~$\tilde{\omega}$.
    (b)~Numerical solution $\tilde{\rho} \equiv
  |\tilde{\psi}(\tilde{t}_j,\, x)|^2$ at fixed time moments
  $\tilde{t}_j$ approaching ${\tilde{t}_*\equiv 0}$.
  (c)~The same
  solution in the self-similar coordinates versus the asymptotic
  profile $\chi_*(y)$.\label{fig:star}}
\end{figure}

In this Letter we present full field-theoretical study of the
collapse. We reveal for the first time its complete dynamics and show
that it leads to explosions of  the Bose stars.  First, the star Bose 
condensate undergoes ``wave 
collapse''~\cite{Zakharov12,  Nature2001}. Namely, it approaches a 
singular density profile due to  self-similar infall of the ALP into
the star center. Second, multiparticle relativistic interactions in the
dense center produce an outgoing stream of
mildly relativistic ALP. These two processes repeat many times.

%%%%%%%%%%%%%%%%%%%%%%%%%%%%%%%%%%%%%%%%%%
\paragraph{2. Nonrelativistic evolution.} Substituting the 
Ansatz $a/f_a = \sqrt{2} \; \mathrm{Re}\left(\psi \,
\mathrm{e}^{-imt}\right)$ into the classical field equation and making
the standard small-field and small-velocity assumptions, one obtains
the nonrelativistic Gross-Pitaevskii-Poisson
equations~\cite{Chavanis:2011, Eby:2016ab}, 
\begin{align}
  \label{eq:24}
  &  i \partial_t \psi = -\Delta \psi/2m  + m(\Phi  -
g_4^2\, |\psi|^2/8)\, \psi \;,\\
\label{eq:25}
& \Delta \Phi  = 4\pi \rho/M_{pl}^2 \;,
\end{align}
which incorporate gravitational and quartic ALP interactions; 
$\Phi(t,\, \boldsymbol{x})$ and $\rho = m^2 f_a^2 |\psi|^2$ are  
the gravitational potential and ALP mass density, respectively.
Note that all physical constants disappear from Eqs.~(\ref{eq:24}),
(\ref{eq:25}) after coordinate and field rescaling: $t =
\tilde{t}/mv_0^2$, $x = \tilde{x}/mv_0$, $\psi = v_0 \tilde{\psi}/g_4$
and $\Phi = v_0^2 \tilde{\Phi}$, where   $v_0 \equiv f_a/g_4M_{pl} \ll
1$. This gives universal description of nonrelativistic evolutions at
essentially different values of parameters. 

Nonrelativistic Bose stars~\cite{Ruffini:1969qy, Chavanis:2011, Eby:2016ab} are stationary
spherically-symmetric solutions to Eqs.~(\ref{eq:24}), (\ref{eq:25})
of the form $\psi = \psi_s(r)\,\mathrm{e}^{-i\omega t}$, $\Phi =
\Phi_s(r)$, where $\omega$ and $\psi_s$ are the binding energy and
wave function of the condensate particles, $r$ is the radial
coordinate. We numerically find the family of Bose stars
parametrized with $\tilde{\omega} \equiv \omega/(mv_0^2)$; their
mass $\tilde{M}(\tilde{\omega})$ is plotted in~Fig.~\ref{fig:star}a. 

The star mass is maximal at some critical value ${\tilde{\omega} = 
  \tilde{\omega}_{cr}}$:  $M_{cr} \equiv M(\tilde{\omega}_{cr}) 
\approx 10.2 \, f_a M_{pl}/mg_4$. This result was first obtained
in~\cite{Chavanis:2011}. The stars with   
$\tilde{\omega} > \tilde{\omega}_{cr}$ are unstable: they violate the
necessary Vakhitov-Kolokolov criterion  $dM/d\omega > 0$~\cite{vk,
  Zakharov12}. Physically, this instability is
caused by the attractive $g_4$-interaction which wins over quantum
pressure and causes the 
stars to collapse. This reveals the fate of the Bose stars  in the
Universe:  they form, grow overcritical by condensing the ALP,
and finally collapse.    

Now,  consider the last stage of the star lifetime, i.e.\ the
collapse. We compute the critical star profile $\psi =\psi_s(r)$ at
$\tilde{\omega} = \tilde{\omega}_{cr}$, then slightly increase its
mass by transforming $\psi_s(r) \to \gamma \psi_s(\gamma r)$ with
$\gamma=1-10^{-4}$. We numerically evolve the resulting overcritical
configuration~\cite{foot2} by solving
Eqs.~(\ref{eq:24}),~(\ref{eq:25}) in spherical symmetry. The numerical
solution $\tilde{\rho}  = |\tilde{\psi}(\tilde{t}_j,\,
\tilde{r})|^2$ is shown 
in Fig.~\ref{fig:star}b at consecutive non-uniformly spaced time
moments $\tilde{t}_j$. At first, the star is thin and wide, and well
described by the Gaussian profile~\cite{Chavanis:2016, Eby:2016}. Then
the profile changes and a central singularity $|\psi| \propto 
r^{-1}$ at $t = t_*$ appears; we set $t_*=0$ without loss of
generality~\cite{foot5}. Spontaneous appearance of singularities is a
nonlinear wave phenomenon called ``wave  collapse'' in  condensed
matter physics~\cite{Zakharov12, Nature2001}. It is nontrivial and  
always related to strongly singular attractive potentials like
${\Delta U \propto   -|\psi|^2}$ in the last term of Eq.~(\ref{eq:24})
with $|\psi|^2 \propto r^{-2}$. 

%%%%%%%%%%%%%%%%%%%%%%%%%%%%%%%%%
\paragraph{3. Self-similar solution.}  To
describe the wave collapse analytically, we ignore the gravitational
potential $m\Phi \ll \Delta U$ at small $r$. After
that we can use results of
  Ref.~\cite{Zakharov12}. Equation~(\ref{eq:24}) acquires a
continuous two-parametric scaling symmetry ${\psi(t,\, x) \to
  \gamma\, \psi(\gamma^2   t,\,  \gamma x)   \,
  \mathrm{e}^{i\alpha}}$ and 
hence the scale-invariant solution $\psi = (-mt)^{-i\omega_*}  \,  
\chi_*(y) \, /mr g_4$, where $y = r\sqrt{-m/t}$ is the self-similar
coordinate and $\chi_*(y)$ satisfies the equation
\begin{equation}
\label{eq:10}
- 2\omega_* \chi  + iy \partial_y \chi = -\partial_y^2 \chi
- |\chi|^2 \chi/4y^2
\end{equation}
following from Eq.~(\ref{eq:24}). We numerically solve
Eq.~(\ref{eq:10}) with finite-energy boundary  conditions $\chi_*(0) =
0$ and $\chi_* \to \chi_0\, y^{-2i\omega_*}$ as $y\to +\infty$. We
find a unique solution with parameters $\omega_*
\approx 0.54$ and $\chi_0 \approx 2.85$ shown in Fig.~\ref{fig:star}c
by black line with open dots.

Solution of Eq.~(\ref{eq:10}) was extensively discussed in
condensed matter physics, see~\cite{Zakharov12}.  Here we demonstrate
that $\chi_*$ is an attractor of the Bose star collapse. Indeed, the
central part of the rescaled numerical solution  $\tilde{r}
\tilde{\psi}(\tilde{t}_j,\, \tilde{r})$ in Fig.~\ref{fig:star}c 
approaches $\chi_*$ as ${\tilde{t}_j\to \tilde{t}_*}$. At  finite $r$
and $t\to t_*\equiv 0$ one finds $y\to +\infty$. Thus,
\begin{equation}
\label{eq:3}
g_4^{~}\, \psi(t_*,\, r) = \chi_0 \cdot (mr)^{-2i\omega_* - 1}
 + \psi_{reg} (r)\;,
\end{equation}
where we separated the large-$y$ asymptotic of $\chi_*$ and
denoted the periphery part of the solution as $\psi_{reg}(r)$,
cf.~\cite{Zakharov12}. The 
small-$r$ behavior of $\psi(t_*,\, r)$ is given by the first  term in 
Eq.~(\ref{eq:3}) which is {\it universal} i.e.\ does not depend  on the
initial conditions. Departures from the universality  at large $mr$ are
described by $\psi_{reg}(r)$. 

We numerically checked that the wave collapse occurs beyond 
spherical symmetry. To this end we added an axially-symmetric
perturbation $\delta \psi / \psi \sim 10^{-2}$ to the critical star
and evolved the resulting configuration with the axially-symmetric 
code. We obtained the solution approaching the same
spherically-symmetric attractor $\chi_*$ and, eventually, the singular
profile~(\ref{eq:3}).

%%%%%%%%%%%%%%%%%%%%%%%%%%%%%%%%%%%%%%%
\paragraph{4. Relativistic evolution.}
The nonrelativistic equations~(\ref{eq:24}), (\ref{eq:25}) are not
applicable in the vicinity of the central singularity at $t \sim
t_*$. First, the ALP velocities $v \sim  \partial_r \psi /m\psi  
\sim - (mr)^{-1}$ in the configuration (\ref{eq:3}) exceed the speed
of light at $m r \lesssim 1$. Second, the small-field approximation 
$\psi \sim a/f_a \ll 1$ breaks down in the same region, and the
higher-order terms of the potential (\ref{eq:2}) become important. We
expect that the condensate flow down the potential  well should 
stop at $m r\lesssim 1$ and $t\approx t_*$  because the full ALP
potential ${\cal V}$ is bounded from below. Instead, relativistic
collisions in the dense center should produce a stream of relativistic
axions escaping the star. 

We study evolution of the Bose star at $t\gtrsim t_*$ by solving
numerically the spherically-symmetric relativistic field equation,
\begin{equation}
\label{eq:5}
f_a \,\Box a = - (1+2\Phi) \; {\cal V}'(a/f_a)\,,
\end{equation}
with the full scalar potential ${\cal V}$. For definiteness,  in what
follows we consider the potential of the QCD
axions~\cite{diCortona:2015ldu}
\begin{equation}
\label{eq:6}
{\cal V} = m^2 f_a^2\, z^{-1} (1+z)\,
\left(1+z-\sqrt{1 + z^2 + 2z \cos\theta}\right)
\end{equation}
with $z \equiv m_u/m_d \approx 0.56$, although the other bounded
potentials produce similar results. The quartic constant in
Eq.~(\ref{eq:2}) is now fixed: $g_4^2 = (1+z^2-z)/(1+z)^2$.   

Note that Eq.~(\ref{eq:5}) still involves the nonrelativistic
gravitational potential $\Phi$ satisfying Eq.~(\ref{eq:25}) with
$\rho = (\partial_t a)^2/2 + (\partial_r a)^2/2 + {\cal V}$. This is
legitimate: departures from the Newtonian gravity occur only in the
region $mr\lesssim 1$ where it has negligible effect on solution,
see~(\ref{eq:3}).  

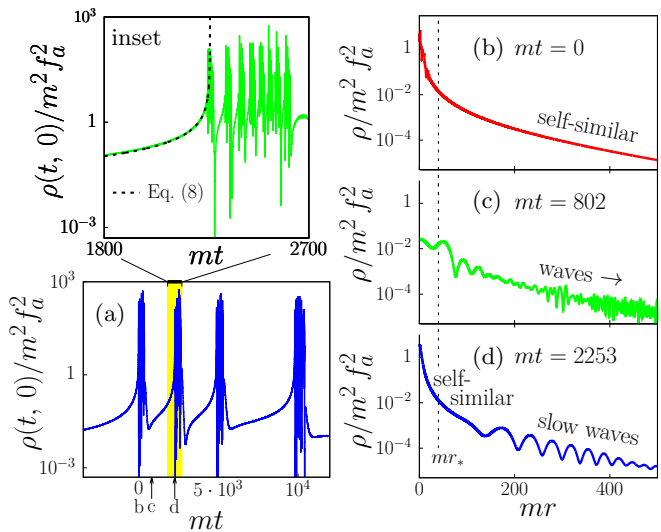
\begin{figure}[t]
  \unitlength=0.01\columnwidth
  \begin{picture}(100,77)(0,-2)
    \put(0,-2){\includegraphics[height=0.8058\columnwidth]{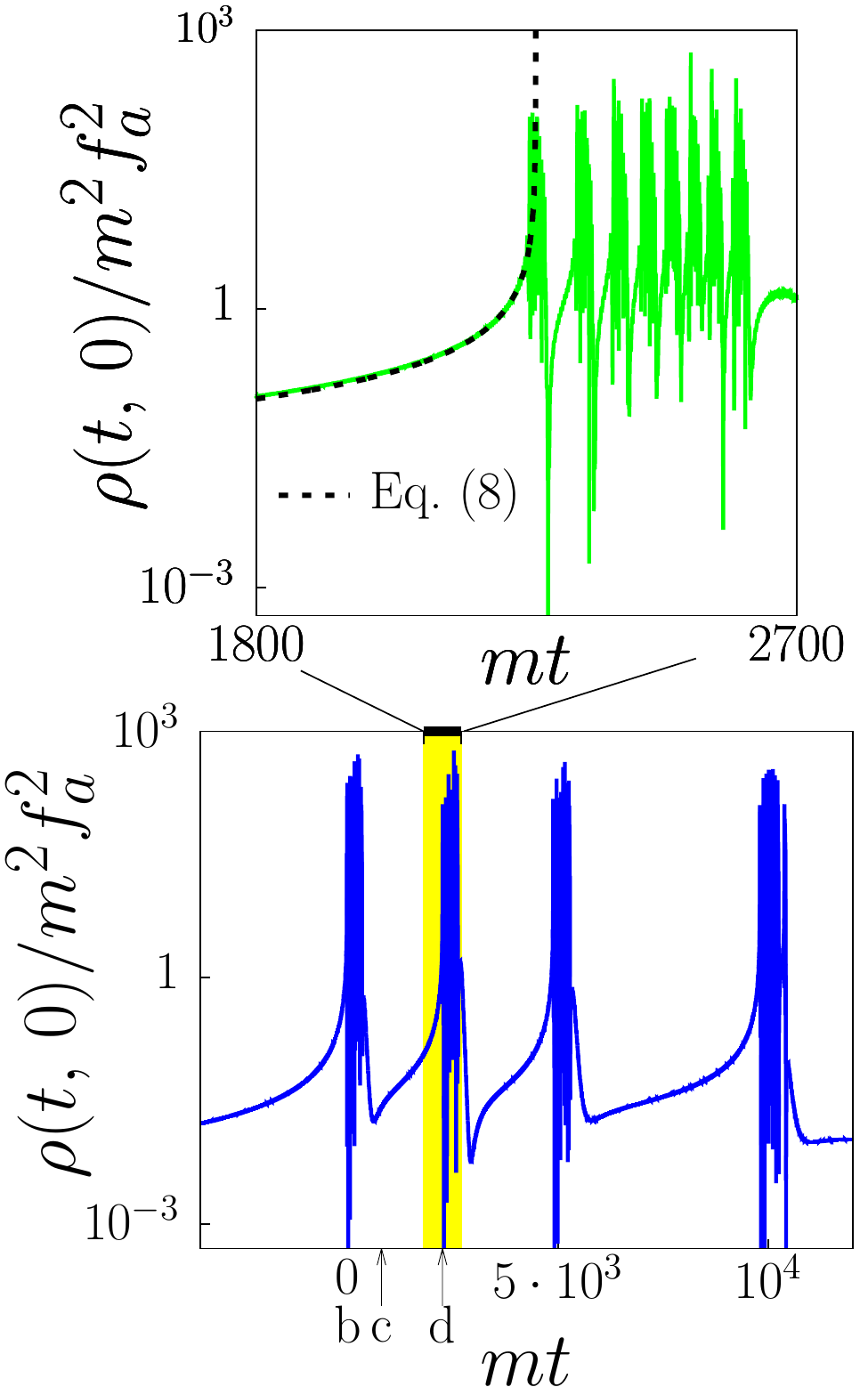}}
    \put(50.7,0){\includegraphics[height=0.775\columnwidth]{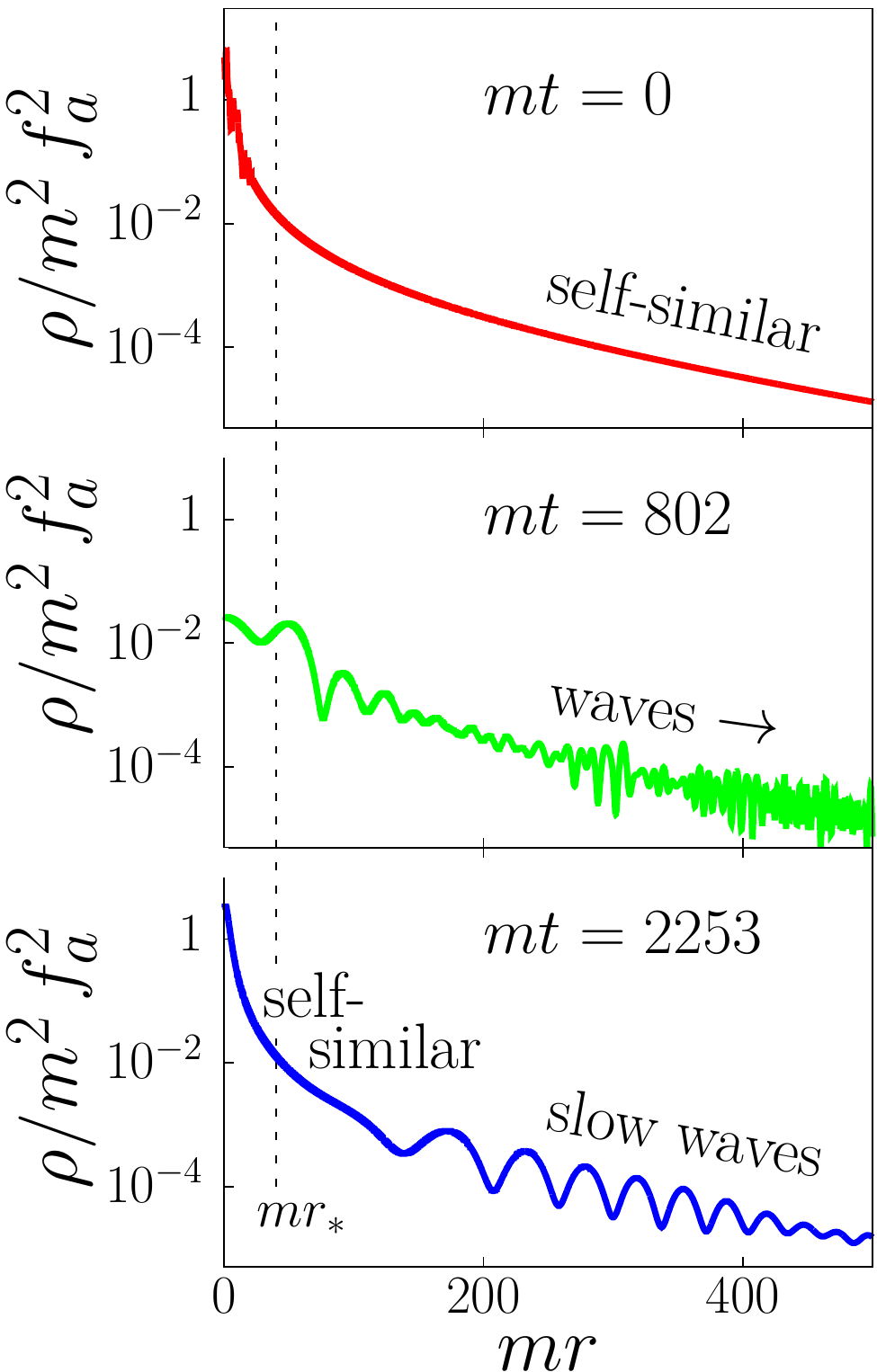}}
    \put(13.4,30){(a)}
    \put(16,72.522){inset}
    \put(71.5,71){(b)}
    \put(71.5,47.5){(c)}
    \put(71.5,23.5){(d)}
    %\put(0,-2){\line(1,0){100}}
    %\put(0,-2){\line(0,1){100}}
    %\put(0,6.1){\line(1,0){100}}
    %\put(0,77){\line(1,0){100}}
    %\put(100,0){\line(0,1){100}}
  \end{picture}
  \caption{(a) The central energy density $\rho(t,\, 0)$ of the
    collapsing star with $f_a^2 = 5\cdot 10^{-8}\, M_{pl}^2$. The 
    region corresponding to the second explosion is magnified in
    the inset. (b), (c), (d)~The density profiles $\rho(t,\, r)$ of this
    star at fixed $t$.\label{fig:2}}
\end{figure}

After finding the numerical solution of Eqs.~(\ref{eq:5}),
(\ref{eq:25}), we plot the energy density in the star center
$\rho(t,\, 0)$ in Fig.~\ref{fig:2}a. One observes that the initial
self-similar growth of  the density at $t<t_*=0$ leads to an
``explosion'' --- a period of violent  oscillations at $t\approx 0$
accompanied by strong emission of the outgoing  high-frequency waves
in Figs.~\ref{fig:2}b,c, see Supplemental Material for the movie. The
oscillations are governed by nonlinear interaction~\cite{foot3} of the
relativistically infalling condensate at $mr \lesssim 1$. In terms of
particle physics this is a relativistic collision of many condensate
particles in the star center producing an outgoing flux of
relativistic ALP. The particle number is strongly  violated in this 
process. 

The outgoing relativistic flux rapidly makes the star center
diluted. Then after the first explosion the process repeats itself:
the central density starts to grow until the second period
of violent oscillations occurs, etc, see Figs.~\ref{fig:2}a,c,d. To
explain this behavior, we recall that the self-similar solution 
studied above is the attractor of the nonrelativistic evolution. This means
that the nonrelativistic condensate outside
of the diluted core should once again develop the  universal singular profile
(\ref{eq:3}) as $t  \to t_*'$. We support this observation by noting
that the central energy density of the self-similar solution $|\psi| =
|\chi_*(y)|/mr g_4$ behaves as
\begin{equation}
  \label{fcenter}
  \rho(t,\, 0)/m^2 f_a^2 = |\partial_y\chi_*(0)|^2 /[g_4^2 m(t_*'-t)]\;,
\end{equation}
with $\partial_y \chi_*(0) \approx 3.99$. In the inset of
Fig.~\ref{fig:2}a we demonstrate that the function $\rho(t,\, 0)$ at
the eve of the second explosion is well fitted by Eq.~\eqref{fcenter}
(shown by the dashed line) with one adjustable parameter $t_*'$. We
conclude that the collapsing star indeed repeatedly develops singular
profiles (\ref{eq:3}) triggering scattering of the relativistic ALP in
the star center.  

%%%%%%%%%%%%%%%%%%%%%%%%%%%%%%%%%%%%%%%%%%%
\paragraph{5. Emission spectrum.}
Since the wave collisions in the star center always start from the
approximately universal initial data~(\ref{eq:3}), their dynamics and
the spectra of the emitted ALP are almost the same. To calculate the 
spectra, we consider the spherical axion waves freely propagating away
from the star at large $r$: $a(t,\, r) \approx
\mathrm{Re}\int_0^{+\infty} dk\, 
A_k\,  \mathrm{e}^{ikr -  i\omega_kt} /r$, where $\omega_k^2=  k^2 +
m^2$. We numerically compute the wave amplitudes $A_k$ using the 
time Fourier transforms of the field $a(t,\, R)$ and its derivative
$\partial_r a(t,\, R)$ at large $r = R$. Then the 
spectrum or, more exactly, the distribution of the total emitted
energy ${\cal E}$ over the momenta $k$ of the outgoing particles, is
${d{\cal   E}/dk = 4 \pi^2\omega_k^2\, |A_k|^2}$.

\begin{figure}[t]
  \unitlength=0.01\columnwidth
  \begin{picture}(100,54.5)
    \put(0,1.2){\includegraphics[height=0.542\columnwidth]{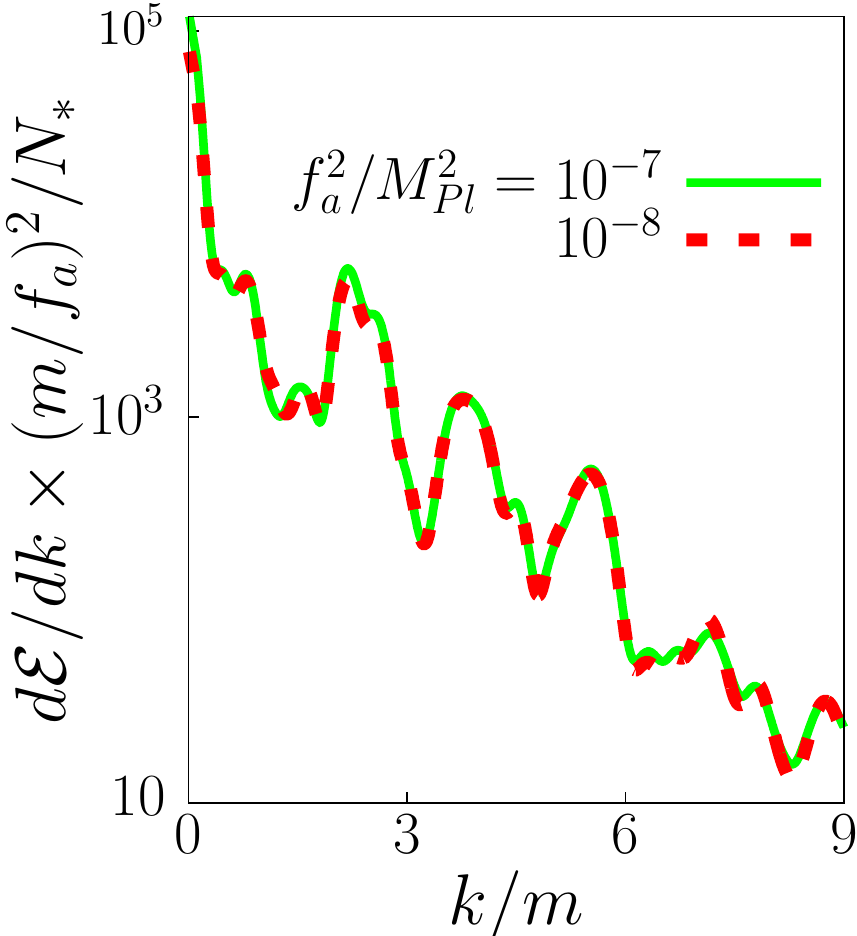}}
    \put(51.1,0){\includegraphics[height=0.56\columnwidth]{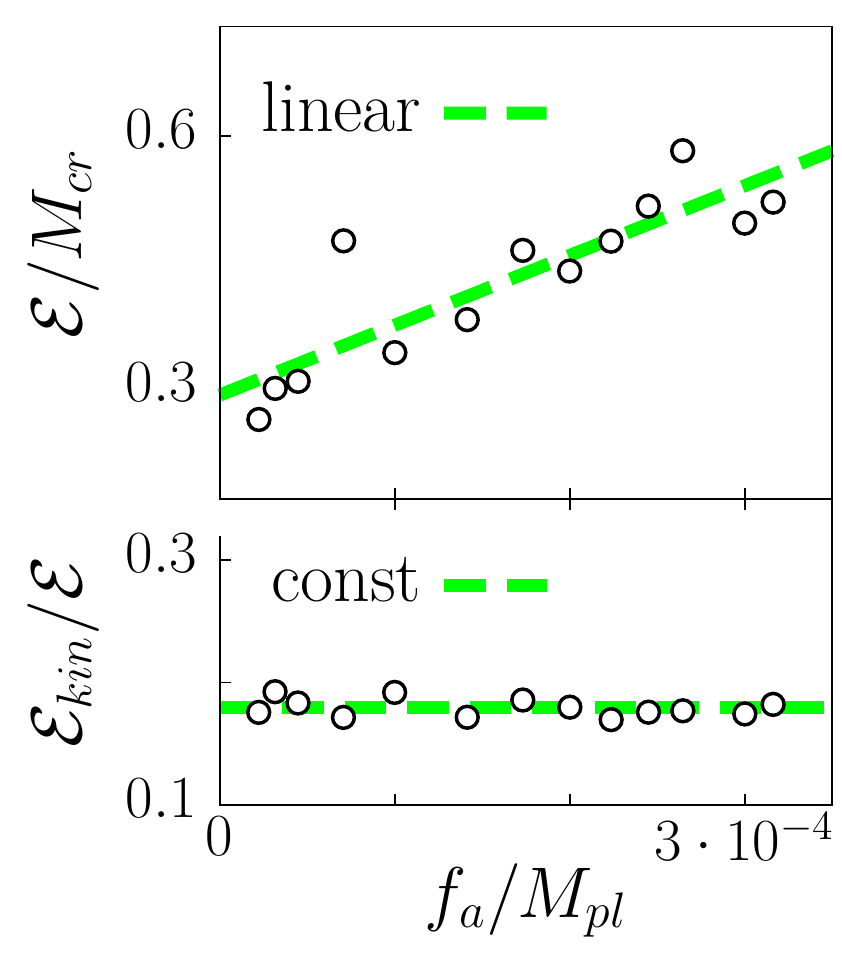}}
    %\put(0,0){\line(1,0){100}}
    %\put(0,8.8){\line(1,0){100}}
    %\put(0,54.5){\line(1,0){100}}
    %\put(100,0){\line(0,1){100}}
    %\put(0,0){\line(0,1){100}}
    \put(42.5,50){(a)}
    \put(93.5,50){(b)}
    \put(93.5,21.5){(c)}
  \end{picture}
  \caption{(a) The spectra $d{\cal E}/dk$ of emitted particles per
    one explosion. The graphs are  averaged over the sensitivity
    intervals $\delta k = 0.1\, m$ for visualization
    purposes. (b), (c) The portion of mass ${\cal E}/M_{cr}$ leaving
    the collapsing star during its full lifetime and the fraction
    of kinetic energy ${\cal E}_{kin}/{\cal E}$ within the outgoing
    particle stream as functions of the ALP scale~$f_a$.
\label{fig:3}}
\end{figure}
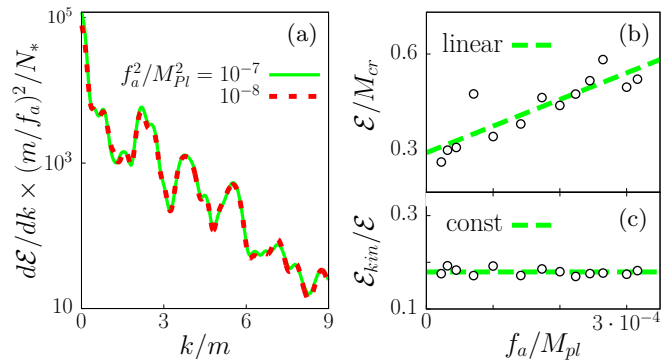

In Fig.~\ref{fig:3}a we plot the spectra emitted by two collapsing stars, each
divided by the number of explosions during the collapse $N_*$. The
spectra are wide, with several almost equidistant peaks~---  these may
appear due to multiple re-scattering of the outgoing ALP off the
infalling condensate~\cite{Khlebnikov:1996mc}. The two graphs in
Fig.~\ref{fig:3}a coincide although the respective processes proceed
at essentially different $f_a$ and involve $N_* = 3$ and $7$
explosions. This confirms that the explosions are identical:  the
total spectrum is given by the one in Fig.~\ref{fig:3}a multiplied by
$N_*$.

However, the number of explosions $N_*$ and therefore the total emitted
energy ${\cal E}$  depend on the potential ${\cal V}$. 
Figure~\ref{fig:3}b shows that ${\cal 
  E}$ is an almost linear function of $f_a/M_{pl}$. Extrapolating it to 
phenomenologically relevant values $f_a \ll M_{pl}$, we find that
the collapsing stars in the model~(\ref{eq:6}) lose roughly $30\%$
of their mass,  ${\cal E}  \sim 0.3 \, M_{cr}$.

%%%%%%%%%%%%%%%%%%%%%%%%%%%%%%%%%%%%%%%%%%%
\paragraph{6. Discussion.}
To summarize, the Bose star collapse involves two alternating
processes: the self-similar evolution approaching the singular
profile~(\ref{eq:3}) and  ALP scattering in the star center
accompanied by strong emission of relativistic particles. The first process 
is described, after the appropriate field and coordinate rescalings, by
the universal nonrelativistic solution. The subsequent scattering,
although non-universal by itself~\cite{foot4}, starts from the universal
profile~(\ref{eq:3}) at each cycle and therefore produces the same
outgoing flux of the axion-like particles. The collapse
stops when the star remnant is not able to support the self-similar
evolution. We numerically checked that the remnant is gravitationally
bound. It may therefore settle into a subcritical star, which, if
surrounded by the ALP minicluster, may again grow overcritical by
condensing the axions and collapse. 

Note that contrary to the recent suggestion~\cite{Chavanis:2016} the
collapsing star does not form a black hole at $f_a\lesssim M_{pl}$
because the central region of the configuration~(\ref{eq:3}) in this
case remains outside of its gravitational radius. On the other hand,
 black hole formation is expected~\cite{Helfer:2016ljl} at
 ${f_a\gtrsim M_{pl}}$. 

Our results essentially modify cosmology and astrophysics of the 
ultralight ALP with $m\sim  10^{-22}\, \mbox{eV}$ and relatively low
ALP scale $f_a < 10^{15} \, \mbox{GeV}$. In this case the largest
masses $M\sim 10^{9}\, M_{\odot}$ of the Bose stars observed in the
full-scale simulation~\cite{Schive:2014dra} are above
$M_{cr}$. The collapse of those stars  into relativistic ALP with
spectrum in Fig.~\ref{fig:3}a should radically change the model
predictions once the self-coupling is taken into account.

Application of our results to other ALP models depends on whether 
their Bose stars populate the Universe or not. In the particular case
of the QCD axions with $m\sim 10^{-4} \, \mbox{eV}$ the value of the
critical mass $M_{cr} \sim 10^{-13} \, M_{\odot}$ is somewhat smaller
than the typical mass of the axion miniclusters. If the conjecture
of~\cite{Kolb:1993zz} is valid and the miniclusters condense into Bose   
stars, the latter should eventually collapse ruining the former and
producing the relativistic axions. This is important for
observations~\cite{Tinyakov:2015cgg}. Another possible application is 
emission, during the collapse, of radio waves which may be directly
observable~\cite{Tkachev:2014dpa}. Their specific
time pattern related to Fig.~\ref{fig:2} and widening of  their
spectrum due to  relativistic axion velocities in the star core,
cf.\ Fig.~\ref{fig:3}a, may serve as distinctive signatures of the
star collapse. 

Prominent cosmological phenomena should appear in very special
ALP dark matter models where Bose stars carry a considerable fraction
of the dark mass at some stage of the Universe evolution,
cf.~\cite{Guth:2014hsa, Davidson:2016uok}. Indeed, the collapse turns
30\% of the total mass of overcritical stars into warm dark matter
with universal spectrum in Fig.~\ref{fig:3}a containing 18\% of kinetic
energy, see Figs.~\ref{fig:3}b,c. The net  effect of this process on
the structure formation and cosmological parameters   should be
similar to that of the dark matter decaying into dark
radiation~\cite{Audren:2014bca}.  

We thank E.~Nugaev and S.~Sibiryakov for
discussions. This work was supported by the grant RSF
16-12-10494. Numerical calculations  were performed on the
Computational cluster of the TD~INR~RAS.

%%%%%%%%%%%%%%%%%%%%%%%%%%%%%%%%%%%%%%%%%


\begin{thebibliography}{99}
\bibitem{foot2} We use nonuniform lattice with adaptive refinement,
  Kreiss-Oliger dissipation, the second-order iterated Crank-Nicholson 
  integrator for the nonrelativistic evolution and the sixth-order 
  Runge-Kutta method for the relativistic one. These advanced
  numerical techniques are targeted at achieving record precision of
  the numerical   solution which starts from $10^{-10}$ in the
  beginning of the collapse and degrades to $10^{-2}$ at $m t\sim
  10^4$.
\bibitem{foot3}
  During these oscillations the field amplitude at $r=0$ exceeds $2\pi
  f_a$ which may be loosely interpreted as a formation of short-living
  axion domain wall surrounding the star center.
\bibitem{foot4}
  The qualitative features of the spectrum, however, are
  model-independent. We confirmed this by performing 
  calculations for   ${\cal V} = m^2 
  f_a^2(1-\cos\theta)$ and the potential with   $\theta^6/6!$ in place
  of dots in Eq.~(\ref{eq:2}). In both cases the spectrum was
  mildly relativistic with several wide peaks.
\end{thebibliography}
\end{document}